# Shacl4Bib—custom validation of library data

Péter Király
Gesellschaft für wissenschaftliche Datenverarbeitung mbH Göttingen (GWDG), Göttingen, Germany. ORCID: 0000-0002-8749-4597, email: pkiraly@gwdg.de

*abstract*: The Shapes Constraint Language (SHACL) is a formal language for validating RDF graphs against a set of conditions. Following this idea and implementing a subset of the language, the Metadata Quality Assessment Framework provides Shacl4Bib: a mechanism to define SHACL-like rules for data sources in non-RDF based formats, such as XML, CSV and JSON. QA catalogue extends this concept further to MARC21, UNIMARC and PICA data. The criteria can be defined either with YAML or JSON configuration files or with Java code. Libraries can validate their data against criteria expressed in a unified language, that improves the clarity and the reusability of custom validation processes.
keywords: Metadata quality assessment, SHACL, MARC, PICA

## Introduction

The well known bibliographic metadata schemas used by library catalogues (such as MARC21, UNIMARC, PICA) contain thousands of data elements. They are prepared for describing many different types of library objects, and quite flexible in many aspects. Libraries, even the largest ones utilise only a fraction of the data elements, for example the Library of Congress utilises 178 data fields out of 229 the standard defined and 1224 sub-fields out of 2667. On one hand we have an abundance of options, on the other the metadata schemas do not provide a tool to make distinctions among these data elements in terms of importance from the library's point-of-view. Such distinctions exist, see the survey conducted by Emma Both asking libraries in the UK about their opinion about the factors of data elements to make a bibliographic record ‘shelf ready’. [Booth, 2020] There are cataloguing agreements, such as the Program for Cooperative Cataloging[1] that describes requirements against particular kinds of bibliographic descriptions, however these are traditional documents made for human readers, and they do not have publicly available machine readable implementations. This makes validation of records against local needs difficult. How to tell a validation tool to handle a data element as mandatory, or that a value of a given data element should start with a particular string (e.g. the abbreviation of the library's name)? But not only rules are interesting: there might be ‘antipatterns’: some well known data problems that occur from time to time. Rules tell us the good practice we would like to follow, while antipatterns describe the bad practice that we would like to avoid. Internally, libraries and other cultural heritage organisations maintain documents about such rules expressed in natural language.[2] In two research software, Metadata Quality Assessment Framework (MQAF) [MQAF, 2024] and QA catalogue [QA catalogue, 2023] we successfully implemented generic validation, that take care about the rules set found in the metadata standards (e.g. checking if a data element is defined in the schema, is it allowed to have multiple instances, is an encoded value follow the encoding rules), but we had to write custom software code to implement local rules, and this approach doesn't scale well. This problem led us to the research questions:
RQ1: How to transform a custom metadata criteria into executable, machine readable representation?
RQ2: What should the validation result contain and how to display it?

[1] https://www.loc.gov/aba/pcc/

## Custom metadata criteria

In order to implement a validation pipeline, we have to solve the following issues: a) how to retrieve the data element value we would like to validate? b) how to validate values that come from different types of serialisation formats? c) how to validate against unified (not format dependent) criteria?

Shapes Constraint Language (SHACL) [SHACL, 2017] is a domain specific formal language for validating Resource Description Framework (RDF) graphs against a set of conditions (expressed also in RDF). RDF follows the linked data and open world assumption paradigms, so in contrast to relational databases an object might have any kinds of properties, that sets aside the concept of 'records'. It has lots of good consequences, but makes validation quite hard, since we can not know purely from the graph if a given property that comes from a different name space is acceptable in its current place. SHACL (and a parallel approach Shape Expressions, ShEx) tries to solve this situation by introducing a language that enables data managers to set rules against RDF datasets. Some parts of SHACL are closely bound to the semantic web concepts and technologies, but an interesting part, a vocabulary of constraints could be reused in other technological contexts. Following this idea MQAF implemented this subset, providing a mechanism to define SHACL-like rules against data sources in non-RDF based formats, such as XML, CSV and JSON (SHACL validates only RDF graphs). The rules can be defined either with YAML or JSON configuration files or with Java code.

SCHACL uses RDF notation to 'address' the data element. If we would like to apply the rules to non RDF, we should find format specific mechanisms that do this job. There are some other domain specific languages to address and retrieve data elements in a structure: XPath for XML, JSONPath for JSON, and different CVS libraries provide simple access by column names for CSV. For the above mentioned bibliographic formats the situation is a bit different because the metadata schemas themselves are mostly independent of the serialisation format. QA catalogue is built on top of another Java library, Marc4J, that provides an internal object model. When it reads a record from an ISO 2709 or a MARCXML file, the library transforms the structure into this model. QA catalogue improved it in two ways. First, the group of supported serialisation formats are larger (including Alephseq, line separated binary MARC, MARC Line, MARCMaker, PICA plain, and normalised PICA) plus it also maps PICA to this model. Second, and more importantly, it adds a semantic layer to each data element that is defined in the standard. The software maintains an extensible Java model of MARC21 and some of its extensions: a group of data elements that are locally defined within a particular organisation (the German, Finnish, Czech, Belgian and British national libraries among others). Moreover: the software could read Avram, a JSON schema for creating machine readable bibliographic metadata schema [Avram, 2024], so other extensions will come in the future via this schema. These abstraction levels enable to support two bibliographic 'addressing languages': *MARCspec* [MARCspec, 2017] for MARC21 and UNIMARC records, and *PICA Path* [PICA Path, 2014] for PICA records. These languages do the same as XPath for XML: select and retrieve data elements from the record. MQAF supports XPath, JSONPath and simple column selectors, QA catalogue extends this repertoire with MARCspec and PICA Path. Both enable the users to specify a path to fields, subfields and other data elements that are available in the particular metadata standard.

[2] The current research is partly based on this kind of documents we received from the British Library, Royal Library of Belgium, Europeana and the Deutsche Digitale Bibliothek.

The extracted values for each of the data sources then are transformed again to a very simple atomic data structure, which on the basic level mimics RDF: it consists of a raw value, an optional language tag, and an optional URI value, and it might have parent and child elements.

With all these components in the toolbox we can execute a validation process in a uniform way, so we do not need to write e.g. XSLT script for validating MARCXML, and custom code for validating the same record in Alephseql. We haven't found a similarly reusable component for executing the validation, thus we developed a 'SHACL engine'.

Validation requires the specification of what to validate (including the criteria) which is represented by a schema configuration that describes the part of the metadata schema we would like to analyse. It could be recorded in a YAML or JSON configuration file, or in the underlying Java API. Here is a short example (in YAML).

```
format: MARC
fields:
- name: 040$a
  path: 040$a
  rules:
  - id: 040$a.minCount
      minCount: 1
  - id: 040$a.pattern
      pattern: ^BE-KBR00
```

It tests only a single data element: MARC subfield $a of field 040 (the encoded form of the original cataloguing agency that created the record). In YAML the '-' character and indentation denote child elements. `format` represents the format of the input data. It can be XML, JSON, CSV, MARC or PICA. Since it is MARC here, the tool will use the MARCspec language to retrieve data elements. `fields` lists data elements we would like to investigate. There is only one child, so we analyse a single MARC subfield. `name` is how the data element is referenced within the configuration. It could be a machine or a human readable string. `path` is the 'address' of the metadata element expressed in an addressing language such as MARCSpec or PICA Path. `rules` denotes the set of rules---we have two rules here. `id` is the identifier of the rule.

This will be the header of the column in CSV, and it could be referenced elsewhere in the configuration file. `minCount` and `pattern` are constraints, the first specifies the minimum number of instances of the data element in the record, the second contains a regular expression which should match the values of all instances of the data element.

## Constraints

Shacl4Bib supports an important part of SHACL constraints that is applicable to other data types, plus some other constraints that are not part of SHACL, but proved to be needed in different bibliographic assessment use cases. Above the constraints that actually do the heavy lifting of the validation, we added some properties that play roles in the reporting part.

With `minCount` and `maxCount` constraints one can specify how many occurrences of a data element a record can have minimum and maximum. The constraints accept numeric values. It is possible to express with them that a data element is mandatory (minCount=1), or non repeatable (maxCount=1).

The value range constraints (`minExclusive`, `minInclusive`, `maxExclusive`, and `maxInclusive`) specifies a numeric range within which the field's value should remain. A lower and higher bound can be set with Boolean operators. Integers or floating point numbers are accepted. By using the exclusive constraints the data element value should be different (greater or smaller) than the specified value, while by inclusive constraints they could be also equal.

String constraints are applicable for string values. `minLength` and `maxLength` specify the minimum and maximum character length of a data element value; they accept an integer parameter. Empty values, too short and too long values can be detected with them. `minWords` and `maxWords` work similarly, but they check the length in number of words instead of number of characters. `hasValue` checks if the data element value is equal to the provided value. It is useful if the catalogue has a property which should be the same in all records (e.g. the abbreviation of the library name as the cataloguing agency). `in` is similar, but it accepts a list of values and checks if the data element's value is one of them. `pattern`'s parameter is a regular expression checking if the value matches to it or not. The expression can match a part of the data value, so it can be used for checking if it contains a substring. To match the whole value, the expression should start with ‘^’ and end with ‘$’ characters.

Thus far the parameters were scalars, however sometimes we would like to compare two data values in a metadata record, for example if the title and the description of a Dublin Core record are the same (that happens in Europeana's collection). `equals` and `disjoint` checks if the value of the current data element equals or not with another data element, while `lessThan` and `lessThanOrEquals` checks if the current one contains a value that is less than or equals to another one. The developers of SHACL tried to minimise the size of its vocabulary, so they did not add ‘greater than’ and ‘greater than or equals’ constraints that can be expressed with the combination of others (using Boolean expressions). In the example code above the field has a path and a unique name. This later will be accepted as the parameter of these constraints, so it doesn't need to repeat the sometimes complex path.

With logical operators it is possible to build complex rules. Each operator accepts a list of rules; `and` passes if *all* these rules passed, `or` passes if *at least one* of them passed, and `not` passes if *none* of them passed.

The following constraints are not defined in SHACL, but they were needed during the course of bibliographic metadata quality assessment projects. `unique` checks if the value of the field is unique across the dataset. It has prerequisites: the content should be preliminary indexed with Apache Solr, the field should have an 'indexField' property that tells the field name in the index (MQAF and QA catalogue both provide mechanisms to index records). `dependencies` checks if other rules have already been checked and passed. It passes if all these rules have passed or resulted in NA, otherwise fails. It is useful in combination with some other constraint as its preliminary condition: the constraint will be checked only if the rules in the dependency list did not fail. Its parameter is a list of rule identifiers. The identifiers should be valid, and the current rule should take place after the ones from which it depends. `contentType` accepts a list of MIME content types (such as 'image/jpeg', 'image/png'), interprets the value of the data element as a URL, fetches it and extracts the HTTP header's content type, then checks if it is one of those specified. Note: checking URLs is time consuming, and the result is depending on the speed of connection and availability of the targeted resource. `dimension` checks if the data value is a URL referencing an image and the image fits to some dimension constraints (in pixel)---that can be the minimum and maximum size of width ('minWidth', 'maxWidth'), of height ('minHeight', 'maxHeight'), or in case it is not important whether width or height is longer: of shorter ('minShortside', 'maxShortside') or of longer sides ('minLongside', 'maxLongside').

For the practicalities of metadata quality assessment MQAF defines a number of properties that is applicable to each rule. These properties are not constraints, they do not check data, but they are helping the documentation, the workflow, and---after the validation---in the report creation. `id` identifies the rule, which will be reflected in the output, and can be referenced in the dependencies constraint. The identifier might also help in cases when the schema configuration file is transformed from a human readable document such as cataloguing rules: it keeps the linkage between them.{`description` documents what the particular rule is doing. It can be anything reasonable, it does not play a role in the calculation, but the user interface can reuse it and displays the text along with the results. `failureScore`, `successScore` and `naScore` are scores which will be calculated if the validation fails, passes or in case the subject is missing (there is no such data element in the record). With scores the user can make distinction between criteria and antipatterns and the importance of them. If it is `hidden` the rule will be applied, but its output will not be present in the overall output. It can be used together with dependencies to set up compound conditions---when the actual value of the preconditions are not important enough to display. `skip` prevents a particular rule from being part of calculation. This could be useful in the development phase when the user starts to create a complex rule but hasn't yet finished, or when the execution of the rule takes a long time (e.g. checking content type or image dimension), and temporarily she would like to turn it off. `debug` enables recording some details of the calculation into the log files (typically the rule identifier, the data value and the rule's result).

## Validation use cases

The concept has been tested on two distinct kinds of data sources. The first of them is the records ingested by the Deutsche Digitale Bibliothek[3] (DDB). DDB is a German cultural heritage portal, with the same roles as Europeana for the EU: it collects metadata records from about a thousand German LAM organisations, and builds a search and discovery interface. In cooperation between DDB and GWDG we set up a quality assessment pipeline that checks the incoming records.[4] The service accepts XML records in different metadata schemas: MARC, LIDO, METS-MODS, and the DDB specific variations of Europeana Data Model and Dublin Core. The metadata experts of DDB defined a set of quality criteria for mandatory data elements of the records in natural language, e.g. ‘There must be an ID for the data record in the supplied data record.’ These criteria are as independent from concrete schemas as possible (there were slight differences where the schema itself does not support a given feature, such as MARC records don't have thumbnails). These sentences have been translated to Shacl4Bib rules manually. As all records are in XML, we use XPath as the ‘addressing language’, and since the structure of these metadata schemas are quite different, we created a distinct schema configuration file for each schema. The main difference between them are the XPath expressions, but the criteria remained almost the same. We reused the original sentence as the description of the rule, and used the same identifiers in the project's documents, in the schema files, and in the user interface as our main references. We also assigned scores to the criteria reflecting their importance. The translation was an iterative process, we tested several options while we found the current version---that is true both for the rule sets, and the sentences that became more precise and specific. The validation had several steps. First the records have been preprocessed: as ‘unique’ constraint requires a Solr index, we indexed the necessary fields. The main validation created CSV files recording the output of the validation of each records, and finally we aggregated these results and calculated statistics for the whole datasets, per metadata schemas, per data provider and per data sets (a unit of harvest) and the combinations of these (the user can select which overview she would like to see). Each overview has two different views: one that follows the order and grouping of the original sentences, and another that grouped the sentences into the FAIR principles (such as internal identifiers supports mainly Accessibility, while external identifiers supports Interoperability aspects). With the scoring we made an extra grouping of the records into the categories of ‘Blocked’ (records have serious issues that the data provider must fix), ‘To be improved’, ‘Acceptable’ and ‘Good’. The user can even check individual records, how they pass the criteria.

The second test was run on MARC21 records of the Royal Library of Belgium and the British Library. The development happened in QA catalogue, and it was able to build upon the already existing support of MARCspec, Solr indexing and other functionalities of both this tool and MQAF. Here the main task was to glue these parts together, and extend the user interface with the custom validation display (the tool already has a validation display against the criteria found in MARC21, PICA, and UNIMARC standards). The process became integral part of QA catalogue's service portfolio, however it turned out, that some criteria that seemed to be trivial at first sight (such as 'subfield $x should be mandatory if field XXX is present') is not possible to implement with the combination of current constraints, new constructions (e.g. an if-else statement) should be created in the future.

[3] https://www.deutsche-digitale-bibliothek.de/

[4] Backend and the schema configuration files: https://github.com/pkiraly/metadata-qa-ddb, user interface code: https://github.com/pkiraly/metadata-qa-ddb-web.

Conclusion: the basic hypothesis that SHACL provides a useful vocabulary in cultural context outside of the semantic web technology proved to be true. To make it an integral part of a validation pipeline a new SHACL engine had to be developed that accepts different 'addressing languages'. Beyond SHACL's core other constraints and properties were needed to support special needs of the cultural heritage domain. It also has been proved however that it is not possible to express all the validation criteria with the current set of constraints.

## Future plans

The concept of Shacl4Bib could be extended in two directions: expanding the vocabulary and making it more user friendly. There are other similar vocabularies that define constraints, such as ShEx in the linked data domain, and---in the world of more traditional data---relational database, XML and JSON schema definition languages. We would like to survey if there are parts of them which are missing from our approach. It already turned out that the vocabularies of metadata experts in different LAM sectors are different from the Shacl4Bib vocabulary. Should we add for example aliases for concepts that are the same, but run under different names? Or should we create aliases for concepts that don't have equivalents in our vocabulary, but could be built with a combination of multiple constraints? The second direction is to enable metadata experts to translate natural language sentences into machine readable configuration files. Together with University of Marburg and Göttingen State and University Library GWDG won a research grant from Deutsche Forschunggemeinschaft for the project 'Agile quality assurance of metadata on cultural objects in the context of data integration processes' (Aqinda). In this project we distil templates for typical natural language sentences expressing criteria, where the user (a metadata expert) can fill gaps in the sentences. The user interface helps to build the path, to specify the parameter values, and it guides the metadata expert through the full workflow of quality assessment including the validation and reporting.

**Acknowledgments.** I would like to thank to Jean Michel Nzi Mba (University of Göttingen), who wrote his bachelor thesis on the topic; to the team of DDB: Francesca Schulze, Cosima Berta, Stefanie Rühle, Claudia Effenberger, Letitia-Venetia Mölck who worked together the implementation for XML based metadata schemas; to Juliane Stiller for contributions to the user interface design; to Jakob Voß (GBV) for the PICA based implementation; to Alan Danskin (British Library) and Hannes Lowagie (Royal Library of Belgium) for sharing documentation about their libraries' custom criteria; to the Aqinda team: Prof. Gabriele Taentzer, Regine Stein, Markus Matoni, Arno Kesper, Lukas Sebastian Hofmann; and finally the members of Europeana Data Quality Committee for discussions and feedback for different aspects of this development.